\documentclass[fleqn,twoside,twocolumn,nofootinbib,showkeys]{revtex4} 
\usepackage[nocpr]{ujp} 

\begin{document}
\title[Formation of the Liquid-Crystalline Phase ]
{FORMATION OF THE LIQUID-CRYSTALLINE\\ PHASE IN POLY(DI-n-HEXYLSILANE)}
\author{N.I. Ostapenko}
\affiliation{Institute of Physics, Nat. Acad. of Sci. of
Ukraine}
\address{46, Nauky Prosp., Kyiv 03028, Ukraine}
\email{ostap@iop.kiev.ua}
\author{Y.V.~Ostapenko}%
\affiliation{Institute of Physics, Nat. Acad. of Sci. of
Ukraine}%
\address{46, Nauky Prosp., Kyiv 03028, Ukraine}%
\author{O.A.~Kerita\,}
\affiliation{Institute of Physics, Nat. Acad. of Sci. of
Ukraine}
\address{46, Nauky Prosp., Kyiv 03028, Ukraine}
\affiliation{National Technical University of Ukraine ``Kiev Politechnic Institute''}
\address{37, Prosp. Peremogy, Kyiv 03056, Ukraine}
\udk{???} \pacs{61.30.Vx. 78.40.Me,\\[-3pt] 78.47.jd} \razd{}

\autorcol{N.I.\hspace*{0.7mm}Ostapenko,
Y.V.\hspace*{0.7mm}Ostapenko, O.A.\hspace*{0.7mm}Kerita}

\setcounter{page}{276}%

\begin{abstract}
We studied the formation of the liquid-crystalline (LC) phase in a
poly(di-n-hexylsilane) (PDHS) film at the heating above the
thermochromic transition temperature and its evolution, when the
samples are cooled to room temperature. For this purpose, we
measured the absorption (293--413~K) and luminescence (5 K) spectra
depending on the annealing temperature, annealing modes, film
thickness, and molecular weight of the polymer. It is shown that the
formation of two LC phases at the heating of the film is associated
with the appearance and a transformation of the positions and the
intensities of two new absorption bands in the region of
{\it gauche}-conformation, as well as the appearance of two bands in the
region of {\it trans}-conformation of the cooled film. It is assumed that
the formation of two LC phases is due to the existence of two
absorption centers, which correspond to different distributions of
the segment lengths in neat polymer. It is shown that, in the
polymers with three different lengths of a Si-backbone (18, 50 and 180
nm), the LC phase reliably appears only in the polymer with the length of
the Si-backbone of about 50 nm. We associate the appearance of new
wide bands in the absorption spectrum of the annealed PDHS film
after the cooling to room temperature with the defect states related
to residual phenomena arising after the transition of the thermally
treated film from the LC state to {\it trans}-conformation.
\end{abstract}
 \maketitle

\section{Introduction}

A lot of photo-physical properties of silicone polymers
(polysilanes) are related to the delocalization of electronic
excitations on the segments of a polymeric chain [1], which consist
of silicon atoms and organic molecules as side groups.\,\,As a
result, the strong absorption in the UV range, strong dependence of
the electron transition energy on the conformation of a polymeric
chain, phenomenon of thermochromism, and high mobility of charge
carriers in these polymers are observed.\,\,The high mobility of
charge carriers determines the use of polysilanes as transport [2]
and luminescence [3] layers in organic electroluminescent devices.
Orienting the polymeric chains increases significantly their
photoconductive characteristics [4], which stimulates the study of
the processes of orientation and organization in these polymers.

Poly(di-n-hexylsilane) (PDHS) is one of the most studied
polysilanes. This polymer is characterized by the order-disorder
thermochromic transition [5] above the phase transition temperature
$T = 315$ K. In the disordered phase, the polymeric chains become
free to move due to the disorientation of side groups [6].
Disordering the polymeric chains results in a significant change in
the absorption spectrum of PDHS films. Namely, the absorption band
of {\it trans}-conformation shifts toward shorter wavelengths by about 50
nm, which indicates the sensitivity of the polymer to large
conformational changes of its fragments and to the orderliness of
its structure.

The study of the diffraction of X-rays and electrons confirmed that
the PDHS polymeric chain has {\it trans}-conformation below the
phase transition temperature and transits in {\it
gauche}-conformation above it, which is accompanied by the
spontaneous appearance of the liquid crystal (LC) phase [7]. The
presence of the LC phase in thermally treated polymeric films is
also confirmed by the optical texture observed by the studies of
PDHS films at $T= 353$~K on a polarizing optical microscope [8--10].
Then the similar results were obtained for a number of polysilanes
[11--13]. The existence of the LC phase was confirmed by a twofold
increase in the birefringence of PDHS films cooled to room
temperature after the heating, as well as by the orientation of the
nematic LC deposited on the PDHS film arising after the heating
[10]. The anomalous thermochromic behavior of the absorption spectra
of poly (di-n-octylsilane) was studied in [14].

In this paper, we used the methods of optical spectroscopy to study
changes in the conformation and the organization of PDHS films,
resulting in their transition from {\it trans}-conformation into
{\it gauche}-conformation, and to form the LC phase by heating the
polymer above the phase transition temperature, as well as the
reverse transition at the cooling of samples down to room
temperature. For this purpose, we measured the absorption (at
293--413~K) and luminescence spectra (at 5~K) depending on the
annealing temperature, annealing modes, film thickness, and
molecular weight of the polymer.

\section{Experimental}

PDHS films were obtained by casting a solution of the polymer in
toluene on a rotating substrate made of fused quartz with the
subsequent drying at room temperature. We investigated the
absorption and luminescence spectra of the PDHS films with different
molecular weights ($M_{wi} = 53600$, 21000, and 219000; $i = 1$--3).
The lenght of the Si-backbones are 18, 15, and 180~nm in this case
corresponding.The absorption spectra were measured on a
spectroscopic-computing complex KSVU-23 in the temperature interval
293--413~K. The heating of samples was carried out in two modes. In
the first case, the sample temperature was changed in this
temperature interval. For this purpose, the tested samples were
placed in an oven located on the optical axis of the device. The
absorption spectra were recorded during the heating, by starting
from room temperature. The sample heated to a certain temperature
was cooled to room temperature, and then the absorption spectrum was
recorded. In the second case, the temperature of a sample was
stabilized at $T= 318$~K, and the spectra were recorded in 10 min
for one hour. Then the sample was cooled to the room temperature,
and its absorption spectrum was recorded.

\begin{figure}%
\vskip1mm
\includegraphics[width=\column]{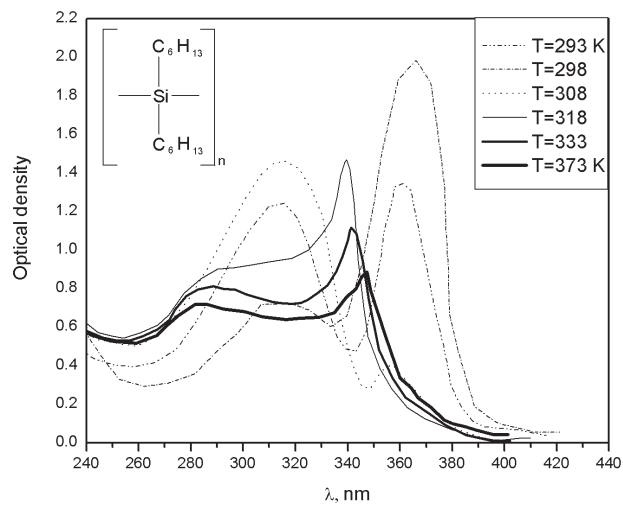}
\vskip-3mm\caption{ Temperature dependence of the absorption spectra
of PDHS films ($M_{w1}$) at the heating in the interval 293--373~K.
Inset shows the PDHS structural formula }
\end{figure}

The luminescence spectra at $T = 5$~K were registered by a
spectrograph DFS-13 combined with an optical helium cryostat. The
luminescence of the samples was excited by light at the wavelength
$\lambda_{\rm ex} = 313$~nm.

\section{Experimental Results}

Figure 1 shows the temperature dependence of the absorption spectra
of PDHS films ($M_{w1}$) under the heating in the temperature
interval 293--373~K. As is known, the absorption spectrum of PDHS
films at 293 K consists of two bands with maxima at 365~nm ({\it
trans}-conformation) and 316~nm ({\it gauche}-conformation) [1]. Let
us consider the dynamics of the band corresponding to {\it
trans}-conformation, as the temperature increases in the interval
293--308~K. In Fig.~1, we can see that, as the temperature
increases, the band shifts to shorter wavelengths. At 308 K, its
maximum is equal to 356 nm. Then, as the film is heated to the phase
transition temperature, it transits from {\it trans}- into {\it
gauche}-conformation. It would be natural to expect that the
absorption spectrum will contain only the {\it gauche}-conformation
band at 316~nm. But we see in Fig.~1 that even a small increase in
the temperature above the phase transition one leads to the
appearance of two new bands instead one band: the narrow band with
maximum at 337~nm and with long-wave wing tightened to 380 nm and
the broad band with maximum at $\sim $287~nm. %
\begin{figure}%
\vskip1mm
\includegraphics[width=\column]{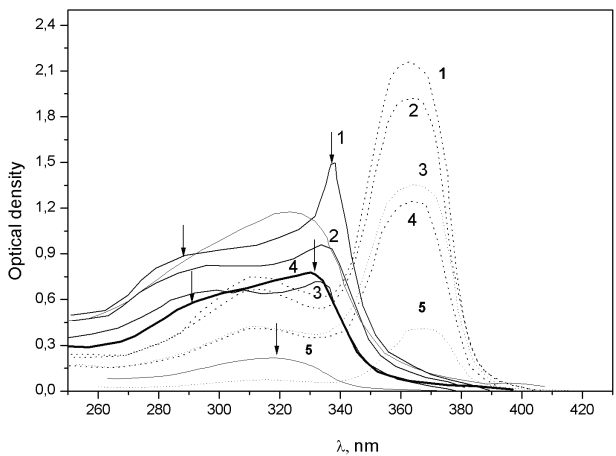}
\vskip-3mm\caption{ Thickness dependence of the absorption spectra
of a PDHS film at 293~K (dotted line) and at 318~K (solid line).
Arrows indicate the maxima of new bands at 337 and 287~nm  with
decreasing the film thickness. Numbers show the spectra of the same
thickness }
\end{figure}%
\begin{figure}%
\vskip3mm
\includegraphics[width=7.2cm]{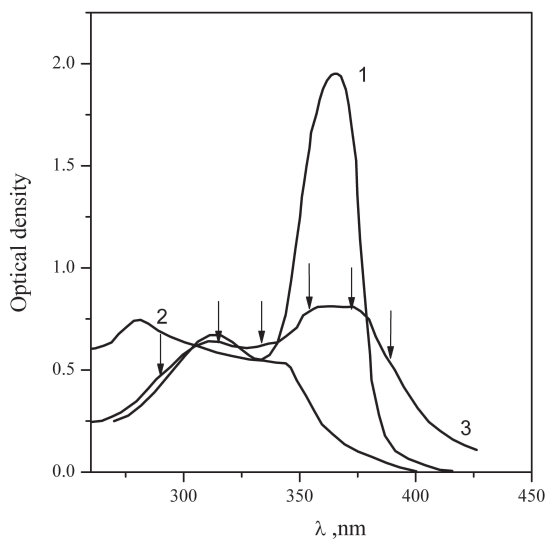}
\vskip-3mm\caption{ Absorption spectra of a PDHS film: {\it 1}~--
as-prepared at 293~K, {\it 2}~-- after the heating to 398~K; {\it
3}~-- after the heating to 398~K and the consequent cooling to
293~K. Arrows indicate the new bands arisen in the absorption
spectra of films cooled to 293~K after the preliminary heating }
\end{figure}%
The further increase in the temperature to 373~K leads to a shift of
these bands by $\sim $7~nm in opposite directions. The bands at 337
and 287~nm are shifted, respectively, to the red and blue sides. As
the temperature increases, the intensities of both bands decrease,
but the intensity of the band at 287~nm decreases less than that of
the 337~nm band. Note that the distance between these new absorption
bands is equal to the distance between the absorption bands of {\it
trans}- and {\it gauche}-conformations in neat polymer at room
temperature.

The appearance of two new absorption bands of {\it
gauche}-conformation, as well as a similar change in their evolution
can be obtained without heating the sample to high temperatures, but
only maintaining its temperature a little higher than the phase
transition temperature (318~K) for about an hour.

We assume that the emergence of two new bands in the absorption
spectrum can be associated with the formation of two LC phases in
the polymeric PDHS film.\,\,The appearance of two new absorption
bands (at 337 and 287~nm), their position, and the intensity of the
absorption spectra of a PDHS film during its heating above the phase
transition temperature strongly depend on the thickness of the
studied film (Fig.~2).\,\,The bands are reliably observed in the
absorption spectrum of the films with optical density $D > 1$
(thickness is about 600~nm) and are not seen in thinner films, e.g.,
in those with optical density $D \sim 0.4$.\,\,As the thickness of
films decreases, the long-wave band maximum is shifted to shorter
wavelengths by about 8~nm (Fig.~2).\,\,The short-wave band maximum
is shifted to longer wavelengths.\,\,However, the intrinsic
mechanisms of the displacements of both bands are different: the
shift of the short-wave band is smaller at large thicknesses of
films and increases, as the film thickness decreases.

At the cooling of a film down to room tempera\-ture for a day after
the preliminary annealing, there are several new bands in the region
of {\it gauche}-con\-formation in the absorption spectrum of films
(Fig.\,\,3): weak bands near 287, 312, and 343~nm.\,\,In the region
of {\it trans}-conformation, we observe two bands with maxima of the
same intensity at 356 and 374~nm instead of a single band with
maximum at 365~nm, as well as the shoulder at 385 nm with a very
lengthy long-wave edge up to 425~nm (Fig.\,\,3, curve {\it
3}).\,\,The significant difference is observed du\-ring the cooling
of samples aged at a temperature above the phase transition
temperature to room temperature.\,\,In this case, the short-wave
band of {\it trans}-conformation is much more intense than the
long-wave one.

\begin{figure}%
\vskip1mm
\includegraphics[width=6.5cm]{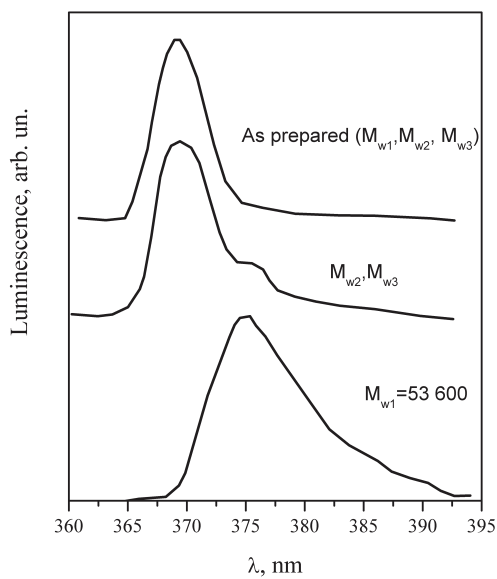}
\vskip-3mm\caption{ Luminescence spectra of as-prepared PDHS films
and thermally treated films with different molecular weights
$M_{wi}$ cooled down to room temperature ($T = 5$~K, $\lambda _{\rm
ex} = 313$~nm) }
\end{figure}

In the luminescence spectrum ($T = 5$~K, $\lambda _{\rm ex} =$ $=
313$~nm) of the PDHS films ($M_{w1}$) cooled after the heating,
there is one band at 375~nm with a wavelength edge tightened to
425~nm. It is shifted relative to the exciton band in the
luminescence spectrum of the original sample to longer wavelengths
by 4~nm (Fig.~4).

The dependence of the absorption spectrum of PDHS films ($M_{w1}$)
on the number of heating cycles is given in Fig.~5. Figure 5 shows
that, as the number of heating cycles increases, the band at 287~nm
in {\it gauche}-conformation becomes more intense, than the band
at 337~nm. As the temperature increases further, their intensities
decrease, and the maxima of both bands are shifted to longer
wavelengths. It should be emphasized that, at the cooling of a
repeatedly heated sample to room temperature, it is clearly seen
that the band of {\it trans}-conformation consists also of two bands
with maxima at 356 and 375~nm. Moreover, the blue band becomes more
intense than the red one, as the number of heating cycles
increases.

\begin{figure}%
\vskip1mm
\includegraphics[width=\column]{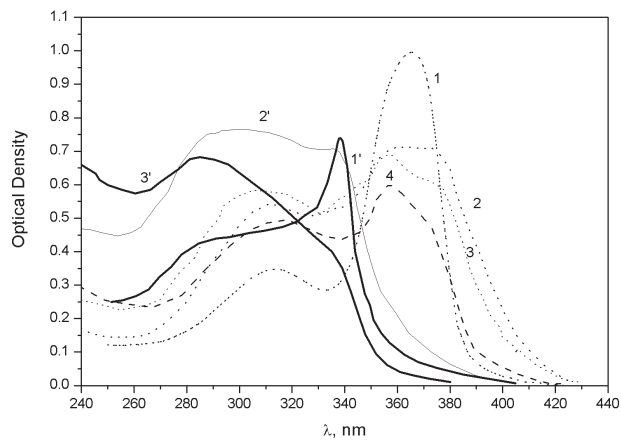}
\vskip-3mm\caption{ Dependence of the absorption spectra of a PDHS
film ($M_{w1}$) on the number of thermal cycles at 318~K ({\it 1, 2,
3, 4}). Dashed lines indicate the absorption spectra: {\it 1}~--
as-prepared film, {\it 2, 3, 4}~-- the film cooled down to room
temperature after the first, second, and third cycles}
\end{figure}

\begin{figure}%
\vskip3mm
\includegraphics[width=\column]{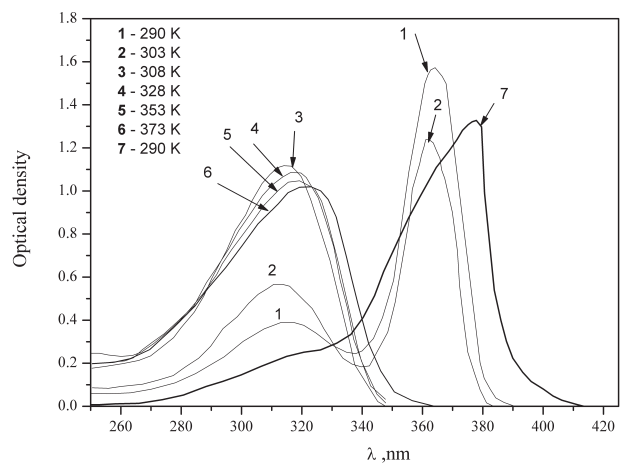}
\vskip-3mm\caption{ Temperature dependence of the PDHS absorption
spectra with $M_{w2}$ at the first heating from 293 to 373~K }
\end{figure}

\begin{figure}%
\vskip1mm
\includegraphics[width=\column]{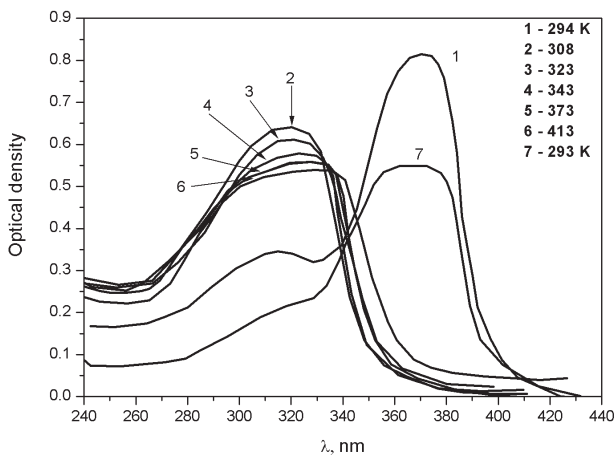}
\vskip-3mm\caption{ Temperature dependence of the PDHS absorption
spectra with $M_{w2}$ at the second thermal cycle }
\end{figure}

\begin{figure}%
\vskip3mm
\includegraphics[width=\column]{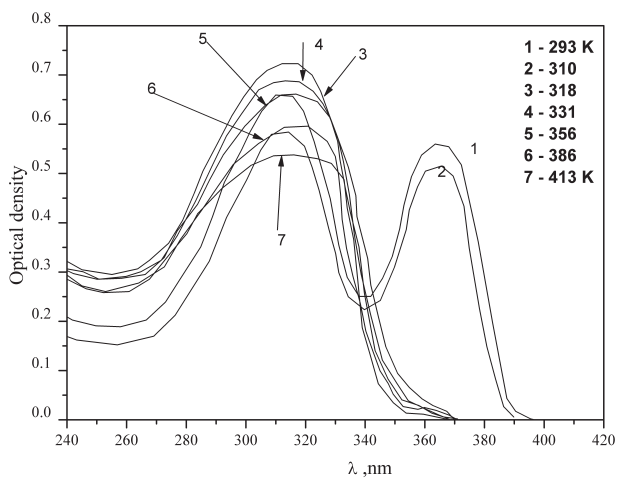}
\vskip-3mm\caption{ Temperature dependence of the PDHS absorption
spectra with $M_{w3}$ at the heating from 293 to 413~K }
\end{figure}

\begin{figure}%
\vskip1mm
\includegraphics[width=5.5cm]{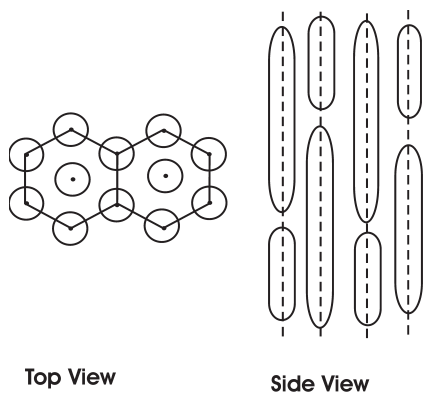}
\vskip-3mm\caption{ Representation of the structure proposed for a
PDHS film after the thermal treatment [7] }
\end{figure}

The studies have shown that the formation of LC phases in the
polymer depends significantly on the molecular weight. The
temperature dependences of the absorption spectra of PDHS films
($M_{w2}$) in the interval from 293 to 373~K in the first and second
heating cycles are shown in Figs.~6 and 7, respectively. It is seen
from Fig.~6 that, in this case, the appearance of LC phases is much
more complicated. With increasing the temperature, we observe only a
slight shift of the 316~nm band to longer wavelengths, which arrives
at 323~nm at 373~K. The second band does not appear at all. The LC
phase appears only after the reheating at a higher temperature $\sim
$343~K (Fig.~7). The temperature dependence of the absorption
spectrum of PDHS films ($M_{w3}$) during the heating from 293 to
413~K is given in Fig.~8. It can be seen that the appearance of LC
phases becomes complicated, and they are observed only at a higher
temperature ($\sim $343~K) than that for the polymer with $M_{w1}$.
In the luminescence spectrum ($T = 5$~K, $\lambda _{\rm ex} =
313$~nm) of PDHS films ($M_{w2}$ and $M_{w3}$) cooled after the
heating, a weak long-wavelength band (Fig.~4) appears at $\lambda =
375$~nm in addition to the exciton band in the spectrum of the
original sample at 371~nm.

\section{Discussion of Results}

We associate the appearance of two new absorption bands at 337~nm
and 287~nm in the region of the band of {\it gauche}-conformation of
the PDHS polymer with the formation of two LC phases at the heating
of the film above the thermochromic transition temperature (Fig. 1).
The spontaneous formation of LC phases in the PDHS polymer at these
temperatures was indicated by studies of the diffraction of X-rays
and electrons [7]. It was shown that the polymeric chains are
oriented in this case in the form of cylinders in the hexagonal
lattice (Fig.~9). The existence of the LC phase in PDHS was also
confirmed by the observation of an optical texture on the surface of
an annealed film, which is characteristic of the classical ordered
structure of liquid crystals [8--10].

We may assume that the intramolecular disordering of side groups
results in the allowedness of the motions of polymeric chains
leading to an increase of the intermolecular interaction between
adjacent polymeric chains. This can cause an enhancement of the
orientation degree of adjacent polymeric chains and their ordering
in the form of cylinders in the hexagonal lattice (Fig.~9). Thus,
the heating of a polymeric film above the phase transition
temperature initiates the spontaneous transition of the polymer in
the LC phase. The LC phases appear, when the polymer is in {\it
gauche}-conformation. Therefore, the formation of these phases and
their variation with the temperature can be observed by the
appearance of new specific features in their absorption spectra,
namely in the region of the band of {\it gauche}-conformation. The
study of the absorption spectra of PDHS films at the heating in the
temperature interval 293--373~K confirms this assumption (Fig.~1).
Indeed, Fig.~1 demonstrates the appearance of two new bands in the
region of {\it gauche}-conformation in the absorption spectrum of a
heated sample instead of a single band at 316~nm in the absorption
spectrum of the initial sample at room temperature. The co-existence
of these new bands in the absorption spectrum of PDHS at 337 and
287~nm up to a temperature of 423~K agrees with the birefringence
revealed by PDHS films at the heating of the polymer above the phase
transition temperature, i.e., in the temperature interval 315--523~K
[6]. It was shown that the birefringence decreases at the heating of
the PDHS film, but does not vanish up to very high temperatures.

It is essential that the new bands shift to opposite sides.\,\,The
displacement of the long-wave band (337~nm) to the long-wave side
testifies, apparently, to a decrease of the distance between
adjacent polymeric chains due to an increase of the intermolecular
interaction between them.\,\,This result can be related, apparently,
to an increase of the arrangement density of polymeric chains, which
coincides with the data obtained in studies of the diffraction of
X-rays.\,\,Since the new bands in {\it gauche}-conformation shift in
different sides, we may assume that the LC phases are located at
different distances from the substrate surface and are related to
two types of ordering of polymeric chains with different lengths of
segments (short and long), respectively.\,\,It should be emphasized
that the study of optical spectra testifies to the appearance of two
LC phases in the polymer, as distinct from the results of studies of
the diffraction of X-rays [7], where a single LC phase was observed.

It is known that the absorption spectrum of the polymer is a
superposition of segments with various lengths.\,\,The study of the
absorption spectrum of the polymer at a low temperature ($T= 2$~K)
showed that the absorption band at 365~nm ({\it trans}-conformation)
become asymmetric, as the temperature decreases [15].\,\,It was
assumed that it consists of two bands (361 and 366 nm) that
correspond to two different spatially separated absorption centers
which have different distributions of the segment lengths of the
polymer chain, i.e., with different collections of short and long
segments, respectively.\,\,The presence of two types of absorptive
packings can be traced at the heating of the film in the interval
293--308~K (Fig.~1).\,\,This is indicated by a shift of the band of
{\it trans}-conformation at 365~nm to the short-wave side, as the
temperature grows, to 356~nm at $T = 308$~K.\,\,Indeed, with
increasing the temperature, the relatively long segments are
disordered in the first place.\,\,Therefore, the absorption band at
365 nm that corresponds to the absorption of shorter segments is
clearly seen.\,\,We assume that these two states in {\it
trans}-conformation are the reason for the formation of two LC
phases.\,\,The formation of these phases from relatively long and
short segments of polymeric chains is supported by the fact that the
distance between the bands at 287 and 337~nm is equal to the
distance between the absorption bands of {\it trans}- and {\it
gauche}-conformations in the spectrum of as-prepared films.\,\,The
contribution of each of these LC phases can be changed, by varying
the number of thermal cycles and the thickness \mbox{of films.}

As the number of heating cycles increases, the short-wave bands in the
region of {\it gauche}- and {\it trans}-conformations become more
intense than the corresponding long-wave bands (Fig.~5). This
testifies to an increase of the degree of energy disordering of the
polymer after the repeated heating.

It is known that if the film thickness decreases to less than 20 nm,
the polymer Si-backbone disorders, and the extensive organization of
polymeric chains is hindered [16]. This implies that the formation
of LC phases must become complicated with decreasing the film
thickness, which is observed in experiments (Fig.~2)

If the appearance of two new absorption bands in an annealed sample
is related to two types of centers with different distributions of
segment lengths of a polymeric chain, then, at its cooling, we would
expect the presence of two bands instead of the single one of {\it
trans}-conformation at 365~nm. It follows from Fig. 3 that, indeed,
the absorption spectrum of the cooled sample contains two bands at
356 and 375~nm instead of the single one of
\textit{trans}-conformation at 365~nm.

At the formation of an LC phase, the polymeric chains orient
relative to one another, which leads, apparently, to an increase of
the concentration of long segments of a polymeric chain. Hence,
while the polymeric PDHS film is cooled down to room temperature
after the heating above the phase transition temperature, the length
of segments of a polymeric chain should increase. Respectively, the
excitation delocalization region should spread. This must cause a
shift of the long-wave absorption band of {\it trans}-conformation
of an annealed film to the long-wave side relative to the initial
position. In the absorption spectrum of a film cooled down to room
temperature after the heating (Fig.~3), we observe, indeed, the
appearance of a band with maximum at 375~nm. This band is shifted by
10~nm to the long-wave side relative to the band in the absorption
spectrum of the initial film, whose position and intensity depend on
the annealing temperature and the \mbox{annealing modes.}

The additional conformation of the presence of oriented structures
in {\it trans}-conformation of PDHS after its preliminary annealing
is given by the appearance of the long-wave band in the luminescence
spectrum of such sample ($M_{w1}$, \mbox{$T = 5$~K}), which is
shifted to the long-wave side by 4 nm relative to the exciton band
in the luminescence spectrum of the initial sample (Fig.~4).\,\,A
significant broadening of this luminescence band as compared with
the exciton band confirms the existence of defects in this polymer
(Fig.~4).\,\,Indeed, as the temperature of the film decreases down
to room temperature, the defect states related to the residual
phenomena at the transformation of the polymer from the LC phase
into {\it trans}-conformation must appear in the polymer.\,\,The
bands at 287 and 343~nm in the region of {\it gauche}-conformation
and the shoulder at 385~nm in the region of {\it trans}-conformation
are related, apparently, to transitions in such defect states
(Fig.~3).\,\,These defect states are related, apparently, to the
formation of clusters in the polymer due to the intermolecular
interaction of adjacent
polymeric \mbox{chains.}

The studies showed that the formation of LC phases in a polymer
depends essentially on its molecular weight and, respectively, the
length of the Si-backbone. The studies of the adsorption (Fig.~1)
and luminescence (Fig.~4) spectra indicate that the LC phases are
formed most reliably in the polymer with $M_{w2}$ (the Si-backbone
length is about 50~nm). In the polymers with less, $M_{w2}$, and
large, $M_{w3}$, molecular weights, the formation of LC phases is
hampered (Figs.~4, 6, and 8). These phases arise only at the
repeated heating of films (Fig.~7) or at a higher temperature
(Fig.~8). Since the formation of LC phases occurs at the heating
above the phase transition temperature as a result of the
orientation and the ordering of polymeric chains, the length of
polymeric chains is significant for a reliable manifestation of
these processes. The polymers with long Si-backbone (180~nm)
require, apparently, higher temperatures for their motion and
orientation. In the case of polymers with small lengths of a
Si-backbone (18~nm), their organization due to a rather free motion
is too complicated, which is observed in
\mbox{experiments.}

\section{Conclusion}

By methods of optical spectroscopy, we have studied the process of
formation of two LC phases in PDHS films, which was initiated by
their heating above the thermochromic transition temperature. The
manifestation of two LC phases is connected with the observation of
two new bands in the region of the absorption band of {\it
gauche}-conformation. It is assumed that the formation of two
absorption centers corresponds to the different distributions of
segments in a neat sample. The formation of LC phases is also related
to the appearance of defect bands in the absorption spectra of
annealed samples cooled down to room temperature. The defect states
are, apparently, cluster structures that are formed as a result of
the intermolecular interaction of adjacent polymeric
\mbox{chains.}\looseness=1

The orientation of polymer chains arisen at the formation of the LC
phase is related to the shifts of the absorption band of
 {\it trans}-conformation and the appropriate luminescence band in
the spectra of annealed films to longer wavelengths.

\rezume{%
Н.І.\,Остапенко, Ю.В.\,Остапенко,
О.О.\,Керіта\vspace*{1mm}}{ФОРМУВАННЯ РІДКО-КРИСТАЛІЧНОЇ\\ ФАЗИ В
ПОЛІ(Ді-н-ГЕКСИЛСИЛАНІ)\vspace*{1mm}} {Вивчено формування
рідко-кристалічної (РК) фази в плівках полі(ді-н-гексилсилані)
(ПДГС) при температурі вище температури термохромного переходу, а
також її еволюцію при охолодженні зразків до кімнатної температури.
З цією метою досліджені спектри поглинання (293--413~K) і
люмінесценції (5~K) залежно від температури та режимів відпалу,
товщини плівки і молекулярної маси полімеру. Показано, що формування
двох РК фаз при нагріванні плівки пов'язано з появою і
трансформацією положення і інтенсивностей двох нових смуг поглинання
в області гош-конформації, а також з появою двох смуг в області
транс-конформації охолодженої плівки. Передбачається, що формування
двох РК фаз пов'язано з існуванням двох центрів поглинання, які
відповідають різним розподілам сегментів по довжині в полімерному
ланцюгу. Показано, що в полімерах з трьома різними довжинами
полімерного ланцюга (18, 50 і 180~нм) РК фаза надійно
спостерігається тільки в полімері з довжиною Si-ланцюга 50~нм. Ми
пов'язуємо появу нових широких смуг в спектрі поглинання відпаленої
плівки ПДГС з дефектами, які відповідають залишковим станам, що
виникають після переходу термічно обробленої плівки від РК стану до
транс-конформації.}

\end{document}